\documentclass[conference]{IEEEtran}

\usepackage{xspace,amsmath,amssymb,amsfonts,epsfig,subfigure,syntonly}
\usepackage{cite,bm,color,url,textcomp,empheq,boxedminipage}
\usepackage{algorithmicx,algorithm}
\usepackage{epstopdf,makecell}
\usepackage{empheq}
\usepackage{pifont}

\usepackage{graphicx,graphics}  
\usepackage{multirow,multicol}
\usepackage{psfrag}    
\usepackage{stfloats}
\usepackage{url}

\newtheorem{lemma}{\underline{Lemma}}[section]

\newcommand{\mv}[1]{\mbox{\boldmath{$ #1 $}}}

\long\def\symbolfootnote[#1]#2{\begingroup
\def\thefootnote{\fnsymbol{footnote}}
\footnote[#1]{#2}\endgroup}

\psfull

\allowdisplaybreaks[4]

\begin{document}

\title{Joint Computation and Communication Cooperation for Mobile Edge Computing}
\vspace{-0.8cm}
\author{\IEEEauthorblockN{Xiaowen~Cao\IEEEauthorrefmark{1},
Feng~Wang\IEEEauthorrefmark{1},
Jie~Xu\IEEEauthorrefmark{1},
Rui~Zhang\IEEEauthorrefmark{2},
and
Shuguang~Cui\IEEEauthorrefmark{3}}
\IEEEauthorblockA{\IEEEauthorrefmark{1}School of Information Engineering, Guangdong University of Technology, Guangzhou, China}
\IEEEauthorblockA{\IEEEauthorrefmark{2}Department of Electrical and Computer Engineering, National University of Singapore, Singapore}
\IEEEauthorblockA{\IEEEauthorrefmark{3}Department of Electrical and Computer Engineering, University of California, Davis, USA}
Email: caoxwen@outlook.com, \{fengwang13, jiexu\}@gdut.edu.cn, elezhang@nus.edu.sg, sgcui@ucdavis.edu
}

\maketitle

\begin{abstract}
This paper proposes a novel joint computation and communication cooperation approach in mobile edge computing (MEC) systems, which enables user cooperation in both computation and communication for improving the MEC performance. In particular, we consider a basic three-node MEC system that consists of a user node, a helper node, and an access point (AP) node attached with an MEC server. We focus on the user's latency-constrained computation over a finite block, and develop a four-slot protocol for implementing the joint computation and communication cooperation. Under this setup, we jointly optimize the computation and communication resource allocation at both the user and the helper, so as to minimize their total energy consumption subject to the user's computation latency constraint. We provide the optimal solution to this  problem. Numerical results show that the proposed joint cooperation approach significantly improves the computation capacity and the energy efficiency at the user and helper nodes, as compared to other benchmark schemes without such a joint design.
\end{abstract}


\section{Introduction}\label{sec:introduction}
Recent technological advancements have enabled various emerging applications (e.g., augmented reality and autonomous driving) that require intensive and low-latency computation at massive wireless devices. As these devices are generally of small size and thus have limited power supply, how to provide them with enhanced computation capability and low computation latency is one crucial but challenging task to be tackled. Mobile edge computing (MEC) has been recognized as a promising technique to provide cloud-like computing at the edge of radio access networks such as access points (APs) and base stations (BSs). By deploying MEC servers therein, wireless devices can offload part or all of their computation-heavy and latency-sensitive tasks to APs and/or BSs for remote execution \cite{Mach17,Mao17,Bar14}. Depending on whether the computation tasks are partitionable or not, the computation offloading can be generally categorized into two classes, namely {\em binary} and {\em partial} offloading, respectively \cite{Mao17}. In binary offloading, the computation task is not partitionable, and thus should be executed as a whole via either local computing at the device itself or offloading to the MEC server. In partial offloading, the task can be partitioned into two or more independent parts, which can be executed by local computing and offloading, respectively, in parallel.

Based on the binary/partial offloading models, there have been a handful of prior works (see, e.g., \cite{Liu16,Niyato15,Bar14,Huang17,Wang17noma,MHchen16CAP,Huang16,Wang17} and the references therein) investigating the joint computation and communication optimization to improve the performance of MEC.
For example, \cite{Liu16} and \cite{Niyato15} considered power-constrained computation latency minimization problems in a single-user MEC system with dynamic task arrivals and channel fading.  \cite{Huang17,Wang17noma,MHchen16CAP} aimed to minimize the system energy consumption while meeting the users' computation latency requirements in multiuser MEC systems. Furthermore, \cite{Huang16,Wang17} proposed interesting wireless powered MEC systems by integrating the emerging wireless power transfer (WPT) technique into MEC, in order to achieve self-sustainable mobile computing.

Fully reaping the benefit of MEC, however, faces several design challenges. For instance, the computation capability at the MEC server and the communication capability at the AP are generally finite; therefore, when the number of supported users increases, the resources allocated to each user would be limited. Furthermore, the computation offloading in MEC systems critically depends on the wireless channel conditions between the users and the AP; hence, when the devices are located far away from the AP or the wireless channels suffer from deep fading, the benefit of offloading would be compromised. To overcome these issues, we notice that future wireless networks will consist of massive devices (e.g., smartphones, wearable computing devices, and smart sensors), each of which is equipped with certain local computation and communication resources; furthermore, at any time instance, it is highly likely that some devices are in the idle status due to the burst nature of both the computation and communication traffics. As a result, enabling user cooperation among these devices in both computation and communication is an efficient and viable solution to improve the MEC performance, where nearby idle devices can share their computation and communication resources to help enhance active computing users' performance.

In this paper, we investigate a new paradigm of {\it{user cooperation in both computation and communication}} for MEC systems. For the purpose of exposition, we consider a basic three-node system, which consists of a user node, a helper node, and an AP node attached with an MEC server. Suppose that the computation tasks need to be executed within a time block, and partial offloading is implemented for computation tasks. To implement the joint computation and communication cooperation, we divide the block into four slots. In the first slot, the user offloads part of its tasks to the helper, such that the helper can cooperatively compute them on behalf of the user in the remaining time. In the second and third slots, the helper works as a decode-and-forward (DF) relay for cooperative communication, in order to help the user offload some other computation tasks to the AP for remote execution in the fourth slot. Under this setup, we pursue an energy-efficient design to minimize the total energy consumption at both the user and the helper, subject to the user's computation latency constraint. Towards this end, we jointly optimize the allocation of time slots, the partition of the user's computation bits (for its local computing, the helper's cooperative computing, and the AP's remote execution, respectively), the central process unit (CPU) frequencies at both the user and the helper, as well as their transmission powers for offloading. Though this problem is non-convex in general, we transform it into a convex form and use the Lagrange duality method to obtain the optimal solution in a semi-closed form. Numerical results show that the proposed joint computation and communication cooperation approach outperforms alternative benchmark schemes without such a joint design.

Note that there have been some prior works that independently studied the communication cooperation (see, e.g., \cite{Tse04,Sen03,relaychannel,Rui14}) and the computation cooperation \cite{xuchenfog,kaibinhuang17}, respectively. In wireless communication systems, the cooperative communications or relaying techniques have been extensively investigated to increase the data rate and/or improve the transmission reliability \cite{Tse04,Sen03,relaychannel,Rui14}.
In MEC systems, the so-called device-to-device (D2D) fogging \cite{xuchenfog} and peer-to-peer (P2P) cooperative computing \cite{kaibinhuang17} have been proposed to enable the computation cooperation between end users, in which one actively-computing user can employ D2D or P2P communication to offload its computation tasks to the nearby idle user for remote execution.
Different from these prior works with sole communication or computation cooperation, this work is the first attempt to pursue the {\it{joint computation and communication cooperation}} by unifying both for maximizing the MEC performance.

\section{System Model and Problem Formulation}\label{sec:system1}

\begin{figure}
\centering
    \includegraphics[width=9cm]{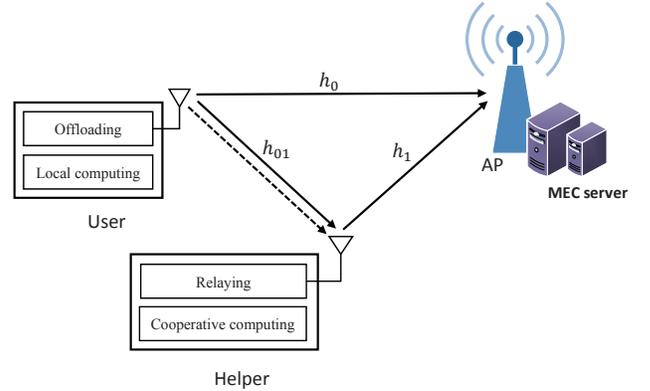}
\caption{A basic three-node MEC system with joint computation and communication cooperation. The dashed and solid lines indicate the task offloading to the helper (for computation cooperation) and to the AP (via the helper's communication cooperation as a relay), respectively.} \label{fig:1}
\end{figure}

As shown in Fig.~\ref{fig:1}, we consider a basic three-node MEC system that consists of one user node, one helper node, and one AP node with an MEC server integrated.\footnote{Note that the joint computation and communication cooperation in this paper can be extended into the scenario with multiple users and multiple helpers, by efficiently pairing one helper with each user for cooperation. The extension is left for our future work.}
All the three nodes are equipped with one single antenna. We focus on a time block with duration $T>0$, where the wireless channels are assumed to remain unchanged over this block, and the user needs to successfully execute computation tasks with $L>0$ input bits before the end of this block. It is assumed that the three nodes perfectly know the global channel state information (CSI) and the computation-related information; accordingly, they can cooperatively schedule their computation and communication resources for the MEC performance optimization.

In order to implement the joint computation and communication cooperation, the $L$ input bits should be generally partitioned into three parts for local computing, offloading to helper, and offloading to AP, respectively. Let $ l_u \ge 0$ denote the number of input bits for local computing at the user, $l_h \ge 0$ denote that for offloading to the helper, and $l_a \ge 0$ denote that for offloading to the AP, respectively. We then have
\begin{align}\label{eqn:L}
l_u+l_h+l_a = L.
\end{align}

\begin{figure}
\centering
 \includegraphics[width=9cm]{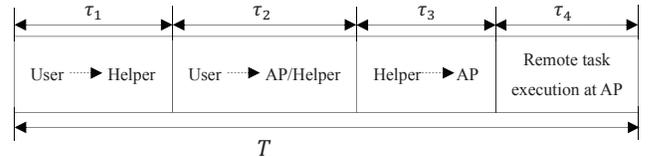}
\caption{MEC protocol with joint computation and communication cooperation.} \label{fig:2}
\end{figure}

\subsection{MEC Protocol With Joint Computation and Communication Cooperation}
In order to implement the joint computation and communication cooperation, the duration-$T$ block is generally divided into four slots as shown in Fig.~\ref{fig:2}. In the first slot with duration $\tau_1\ge 0$, the user offloads the $l_h$ task-input bits to the helper, and the helper can execute them in the remaining time with duration $T - \tau_1$. In the second and third slots, the helper acts as a DF relay to help the user offload $l_a$ task-input bits to the AP. Specifically, in the second slot with duration $\tau_2\ge 0$, the user broadcasts the $l_a$ input bits to both the AP and the helper simultaneously; after successfully decoding, the helper forwards them to the AP in the third slot with duration $\tau_3\ge 0$. After collecting the input bits from the user, the MEC server can remotely execute the offloaded tasks in the fourth time slot with duration $\tau_4\ge 0$.
It is worth noting that we have ignored the time for downloading the computation results from the helper and AP to the user, due to the fact that the computation results are normally with much smaller size than the input bits, and thus the downloading time becomes negligible.
In order to ensure the computation tasks to be successfully executed before the end of this block, we have the following time constraint:
\begin{align}\label{eqn:T}
\tau_1+\tau_{2}+\tau_{3}+\tau_4 \leq T.
\end{align}

In the following, we first introduce the computation offloading from the user to the helper and the AP, and then present the computing at the three nodes.

\subsection{Computation Offloading}
\subsubsection{Computation Offloading to Helper}
In the first slot, the user offloads $l_h$ task-input bits to the helper with the transmit power $P_1\ge 0$. Let $h_{01}>0$ denote the channel power gain from the user to the helper, and $B$ the system bandwidth. Then the achievable data rate (in bits/sec) for offloading from the user to the helper is given by
\begin{align}\label{eqn:r01}
r_{01}(P_1)=B\log_2\left(1+\frac{P_1h_{01}}{\Gamma\sigma_1^2}\right),
\end{align}
where $\sigma_1^2$ represents the power of the additive white Gaussian noise (AWGN) at the receiver of the helper, and $\Gamma\ge1$ is a constant term accounting for the gap from the channel capacity due to a practical modulation and coding scheme. For simplicity, $\Gamma=1$ is assumed throughout this paper. In practice, the number of offloaded bits $l_h$ from the user to the helper cannot exceed $\tau_1r_{01}(P_1)$. Hence, we have
\begin{align}\label{eqn:l1}
l_h \leq \tau_1r_{01}(P_1).
\end{align}
Furthermore, let $P_{u,\max}$ denote the maximum transmit power at the user, and accordingly we have $
P_1 \le P_{u,\max}$. We consider the user's transmission energy as the sole energy budget for computation offloading, and ignore the energy consumed by circuits in the radio-frequency (RF) chains, baseband signal processing, etc. Therefore, in the first slot, the energy consumption for the user's offloading is given by
\begin{align}\label{eqn:par:E1off}
E_1^{\rm offl}=\tau_1P_1.
\end{align}

\subsubsection{Computation Offloading to AP Assisted by Helper}

In the second and third slots, the helper acts as a DF relay to help the user offload $l_a$ task-input bits to the AP. In the second slot with duration $\tau_2$, let $P_2$ denote the user's transmit power with $0\le P_2\leq P_{u,\max}$. In this case, the achievable data rate from the user to the helper is given by $r_{01}(P_2)$ with $r_{01}(\cdot)$ defined in (\ref{eqn:r01}). Furthermore, by denoting $h_0>0$ as the channel power gain from the user to the AP, the achievable data rate from the user to the AP is
\begin{align}
r_0(P_2) = B\log_2\left(1+\frac{P_2h_{0}}{\sigma_0^2}\right),
\end{align}
where $\sigma_0^2$ is the AWGN power at the AP receiver.

After successfully decoding the received message, the helper forwards them to the AP in the third slot with duration $\tau_3$ by using the transmit power $P_3\ge 0$. Let $P_{h,\max}$ denote the maximum transmit power from the helper, and thus it holds that $P_3 \le P_{h,\max}$. Let $h_{1}>0$ denote the channel power gain from the helper to the AP. The achievable data rate from the helper to the AP is
\begin{align}
r_1(P_3)=B\log_2\left(1+\frac{P_3h_{1}}{\sigma_0^2}\right).
\end{align}

By combining the second and third slots, the maximum number of data bits that can be transmitted from the user to the AP via the DF relay (the helper) is given as \cite{relaychannel} $\min\left(\tau_2r_0(P_2)+\tau_3r_1(P_3),~\tau_2r_{01}(P_2)\right)$, which is the upper bound for the number of the offloaded bits $l_a$ to the AP, i.e.,
\begin{align}\label{eqn:l2}
l_a \leq \min\left(\tau_2r_0(P_2)+\tau_3r_1(P_3),~\tau_2r_{01}(P_2)\right).
\end{align}
Similarly as for (\ref{eqn:par:E1off}), we consider the user's and helper's transmission energy consumption for offloading as the energy budget in the second and third slots, respectively, which are expressed as follows.
\begin{align}
E_2^{\rm offl} &=\tau_2P_2 \label{eqn:par:E2off}\\
E_3^{\rm offl} &=\tau_3P_3.\label{eqn:par:E3off}
\end{align}

\subsection{Computing at User, Helper, and AP}

\subsubsection{Local Computing at User}
The user executes the computation tasks with $l_u$ input bits throughout the whole block with duration $T$. Let $c_u$ denote the number of CPU cycles for computing one bit at the user, and $f_{u,n}$ the CPU frequency for the $n$-th CPU cycle, where $n\in\{1,\ldots,c_u l_u\}$. In practice, the CPU frequency $f_{u,n}$ is upper bounded by a maximum value, denoted by $f_{u,\max}$, i.e.,
\begin{equation}\label{eqn:f_u}
f_{u,n}\leq f_{u,\max} .
\end{equation}
As all the local computing should be accomplished before the end of the time block, we have the following computation latency requirement:
\begin{equation}\label{eqn:latency}
\sum_{n=1}^{c_ul_u}\frac{1}{f_{u,n}}\leq T.
 \end{equation}
Accordingly, the user's energy consumption for local computing is \cite{Mao17}
\begin{equation}
E_u^{\rm comp} = \sum_{n=1}^{c_ul_u}\kappa_uf_{u,n}^2,\label{eqn:energy:local:user}
\end{equation}
where $\kappa_u$ denotes the effective capacitance coefficient that depends on the chip architecture at the user. It has been shown in \cite[Lemma~1]{Wang17} that in order for the user to save the computation energy consumption while minimizing the latency, it is optimal to set the CPU frequencies to be identical for different CPU cycles. By using this fact and letting the constraint in (\ref{eqn:latency}) be met with strict equality, we have
\begin{align}\label{eqn:f_u_max}
f_{u,1}=f_{ u,2}=...=f_{ u,c_ul_u}=c_ul_u/T.
\end{align}
Substituting \eqref{eqn:f_u_max} into (\ref{eqn:energy:local:user}), the user's energy consumption for local computing $E_u^{\rm comp}$ is re-expressed as
\begin{align}\label{eqn:par:Eu_comp}
E_u^{\rm comp}=\frac {\kappa_uc_u^3l_u^3}{T^2}.
\end{align}

By combining \eqref{eqn:f_u_max} with the maximum CPU frequency constraint \eqref{eqn:f_u}, we have
\begin{align}\label{eq:f0:max}
c_ul_u\leq Tf_{u,\max}.
\end{align}

\subsubsection{Cooperative Computing at Helper}

After receiving the offloaded $l_h$ task-input bits in the first time slot, the helper executes the tasks during the remaining time with duration $T-\tau_1$. Let $f_{h,n}$ and $f_{h,\max}$ denote the CPU frequency for the $n$-th CPU cycle and the maximum CPU frequency at the helper, respectively. Similarly as for the local computing at the user, we set the helper's CPU frequency for each CPU cycle $n$ as $f_{h,n}=c_hl_h/(T-\tau_1)$, $\forall n \in \{1,\ldots,c_hl_h\}$, where $c_h$ represents the number of CPU cycles for computing one bit at the helper. Accordingly, the energy consumption for the cooperative computation at the helper is
\begin{equation}\label{eqn:par:Eh_comp}
E_h^{\rm comp}=\frac {\kappa_hc_h^3l_h^3}{(T-\tau_1)^2},
\end{equation}
where $\kappa_h$ is the effective capacitance coefficient of the helper.

As in \eqref{eq:f0:max}, we have the following constraint on the offloaded bits due to the maximum CPU frequency $f_{h,\max}$:
\begin{align}\label{eq:f1:max}
c_hl_h\leq (T-\tau_1)f_{h,\max}.
\end{align}

\subsubsection{Remote Computing at AP}
In the fourth slot with duration $\tau_4$, the MEC server at the AP executes the offloaded $l_a$ task-input bits. As the MEC server normally has a stable energy supply (e.g., connected to the grid), the MEC server can compute tasks at its maximal CPU frequency, denoted by $f_{a,\max}$, in order to minimize the computation time. Hence, the time duration $\tau_4$ for the MEC server to execute the $l_a$ offloaded bits is
\begin{align}\label{eq:fa:max}
\tau_4 =l_a/f_{a,\max}.
\end{align}

By substituting \eqref{eq:fa:max} into \eqref{eqn:T}, the time allocation constraint is re-expressed as
\begin{align}\label{eqn:T1}
\tau_1 + \tau_2 + \tau_3 + l_a/f_{a,\max} \le T.
\end{align}

\subsection{Problem Formulation}\label{sec:formulation}
In this work, we aim to minimize the total energy consumption at both the user and the helper (i.e., $\sum_{i=1}^3 E_i^{\rm offl} + E_u^{\rm comp} + E_h^{\rm comp}$ with $E_i^{\rm offl}$'s given in (\ref{eqn:par:E1off}), (\ref{eqn:par:E2off}), and (\ref{eqn:par:E3off}), $E_u^{\rm comp}$ in (\ref{eqn:par:Eu_comp}), and $E_h^{\rm comp}$ in (\ref{eqn:par:Eh_comp})), subject to the user's computation latency constraint, by jointly optimizing their computation and communication resource allocation.{\footnote{As the AP normally has reliable power supply, its energy consumption is not the bottleneck of this MEC system. Therefore, we only focus on minimizing the user's and helper's energy consumption for communication and computation at the wireless devices side. }} The decision variables include the allocation of the time slots ${\bm \tau}=[\tau_1,\tau_2,\tau_3]$, the task partition ${\bm l}=[l_u,l_h,l_a]$, and the offloading power allocations ${\bm P}=[P_1,P_2,P_3]$. Mathematically, the energy-efficient joint computation and communication cooperation problem is formulated as
\begin{subequations}\label{eq.part_prob}
\begin{align}
({\rm P1}):~\min_{{\bm P}, {\bm \tau}, {\bm l}} &~\sum_{i\in\{1,2,3\}}\tau_iP_i+\frac {\kappa_uc_u^3l_u^3}{T^2}+\frac {\kappa_hc_h^3l_h^3}{(T-\tau_1)^2}\label{eq.part_prob.a}\\
{\rm s.t.} &~ 0\leq P_j\leq P_{u,\max},~\forall j\in \{1, 2\} \label{eq.part_prob.b}\\
&~0\leq P_3\leq P_{h,\max}\label{eq.part_prob.b1}\\
& ~0\leq \tau_i\leq T, ~\forall i\in\{1,2,3\} \label{eq.part_prob.c}\\
&~ l_u \ge 0,~ l_h \ge 0,~ l_a \ge 0\label{eq.part_prob.d}\\
&~\eqref{eqn:L},~\eqref{eqn:l1},~ \eqref{eqn:l2},~\eqref{eq:f0:max},~\eqref{eq:f1:max},~{\text{and}}~\eqref{eqn:T1}.   \nonumber
\end{align}
\end{subequations}
In problem (P1), the constraints (\ref{eqn:L}) and (\ref{eqn:T1}) denote the task partition and  time allocation constraints; (\ref{eqn:l1}) and (\ref{eqn:l2}) ensure that the numbers of the offloaded bits from the user to the helper and the AP are limited by the achievable data rates over the respective wireless channels; \eqref{eq:f0:max} and \eqref{eq:f1:max} correspond to the maximum CPU frequency constraints at the user and the helper, respectively.
Note that problem (P1) is non-convex in general due to the coupling of $\tau_i$ and $P_i$ in the objective function \eqref{eq.part_prob.a} and the constraints \eqref{eqn:l1} and \eqref{eqn:l2}. However, we next transform it into a convex problem and solve it optimally in Section~\ref{sec:partial}.

Before solving problem (P1), we first examine the feasibility to check whether the three-node MEC system can support the task execution within the latency constraint.
Towards this end, we obtain the maximum number of supportable task-input bits, denoted by $L_{\max}$. If $L_{\max}$ is no smaller than $L$ in (P1), then (P1) is feasible. Otherwise, (P1) is infeasible. In particular, $L_{\max}$ can be obtained by letting all the three nodes use up all their communication and computation resources, via setting $P_1=P_2= P_{u,\max}, P_3= P_{h,\max}$ and setting the constraints in \eqref{eqn:l2},~\eqref{eq:f0:max},~\eqref{eq:f1:max}, and \eqref{eqn:T1} to be met with strict equality. As a result, we have
\begin{align}\label{eq.part_feaprob1}
L_{\max} \triangleq \max_{{\bm \tau}, {\bm l}}&~l_u+l_h+l_a \\
{\rm s.t.}&~ l_h\leq \tau_1r_{01}(P_{u,\max}),~l_a\leq \tau_2r_{01}(P_{u,\max}) \nonumber\\
&~ c_u l_u=T f_{u,\max},~ c_hl_h=(T-\tau_1)f_{h,\max} \nonumber\\
&~\tau_1 + \tau_2 + \tau_3 + l_a/f_{a,\max} =T \nonumber\\
&~\tau_2r_0(P_{u,\max})+\tau_3r_1(P_{h,\max})=\tau_2r_{01}(P_{u,\max})\nonumber\\
&~\eqref{eq.part_prob.c},~{\text{and}}~\eqref{eq.part_prob.d}.\notag
\end{align}
Note that problem (\ref{eq.part_feaprob1}) is a linear program (LP) and thus can be efficiently solved via standard convex optimization techniques such as the interior point method\cite{Boyd2004}. After $L_{\max}$ obtained, the feasibility of (P1) is efficiently checked. In the next section, we focus on solving (P1) when it is feasible.


\section{Optimal Solution to ({\rm P1})}\label{sec:partial}

This section presents the optimal solution to problem (P1). Towards this end, we first transform it into a convex form by introducing a set of auxiliary variables ${\bm E}=[E_1,E_2,E_3]$ with $E_i \triangleq P_i\tau_i$, where $ i\in\{1,2,3\}$. Accordingly, we have $P_i=E_i/\tau_i$, where we define $P_i=0$ if either $E_i=0$ or $\tau_i=0$ holds, $\forall i\in\{1,2,3\}$. By substituting $P_i=E_i/\tau_i$, $\forall i\in\{1,2,3\}$, problem (P1) is reformulated as
\begin{subequations}\label{eq.part_prob.1}
\begin{align}
({\rm P1.1}):~\min_{{\bm E}, {\bm \tau}, {\bm l}} &\sum_{i\in\{1,2,3\}}E_i+\frac {\kappa_uc_u^3l_u^3}{T^2}+\frac {\kappa_hc_h^3l_h^3}{(T-\tau_1)^2} \label{eq.part_prob.1.a}\\
{\rm s.t.} &~ l_h\leq \tau_1r_{01}\left(\frac{E_1}{\tau_1}\right)\label{eq.part_prob.1.b} \\
&~ l_a \leq \tau_2r_0\left(\frac{E_2}{\tau_2}\right)+\tau_3r_1\left(\frac{E_3}{\tau_3}\right)\label{eq.part_prob.1.c}\\
&~l_a \leq  \tau_2r_{01}\left(\frac{E_2}{\tau_2}\right)\label{eq.part_prob.1.d}\\
&~ 0\leq E_j\leq \tau_jP_{u,\max},~\forall j\in\{1,2\} \label{eq.part_prob.1.e}\\
&~ 0\leq E_3\leq \tau_3P_{h,\max} \label{eq.part_prob.1.e1}\\
&~\eqref{eqn:L},~\eqref{eq:f0:max},~\eqref{eq:f1:max},~\eqref{eqn:T1},~\eqref{eq.part_prob.c},~{\text{and}}~\eqref{eq.part_prob.d},\nonumber
\end{align}
\end{subequations}
where \eqref{eq.part_prob.1.c} and \eqref{eq.part_prob.1.d} follow from \eqref{eqn:l2}. Note that the function $r_j(x)$ is a concave function with respect to $x \ge 0$ for any $j\in\{0,1,01\}$, and therefore, its perspective function $x r_j\left(\frac{y}{x}\right)$ is jointly concave with respect to $x> 0$ and $y\ge 0$. As a result, the constraints (\ref{eq.part_prob.1.b}), (\ref{eq.part_prob.1}c), and (\ref{eq.part_prob.1}d) become convex. Furthermore, the function $l^3/\tau^2$ is jointly convex with respect to $l \ge 0$ and $\tau > 0$, and hence the term $\frac {\kappa_hc^3l_h^3}{(T-\tau_1)^2}$ in the objective function is jointly convex with respect to $l_h\ge 0$ and $0\le \tau_1<T$. Hence, problem (P1.1) is convex and can be efficiently solved by standard convex optimization techniques such as the interior-point method\cite{Boyd2004}. Alternatively, we next use the Lagrange dual method to obtain a well-structured solution for gaining essential engineering insights.

Let $\lambda_1\ge 0$, $\lambda_2\ge 0$, and $\lambda_3\ge 0$ denote the Lagrange multipliers associated with the constraints in (\ref{eq.part_prob.1.b}), (\ref{eq.part_prob.1.c}), and (\ref{eq.part_prob.1.d}), and $\mu_1 \ge 0$ and $ \mu_2$ denote the Lagrange multipliers associated with the constraints in \eqref{eqn:T1} and \eqref{eqn:L}, respectively. For notational convenience, we denote ${\bm \lambda}\triangleq [\lambda_1, \lambda_2, \lambda_3]$ and ${\bm \mu}\triangleq [\mu_1, \mu_2]$. The partial Lagrangian of problem (P1.1) is
\begin{align}
&{{\cal L}}({\bm E},{\bm \tau}, {\bm l}, {\bm \lambda}, {\bm \mu})~\nonumber\\
=&~E_1+\mu_1\tau_1+\frac {\kappa_hc_h^3l_h^3}{(T-\tau_1)^2}+(\lambda_1-\mu_2)l_h+\lambda_1\tau_1r_{01}\left(\frac{E_1}{\tau_1}\right)\nonumber\\
 &+E_2-\lambda_2\tau_2r_0\left(\frac{E_2}{\tau_2}\right)-\lambda_3\tau_2r_{01}\left(\frac{E_2}{\tau_2}\right)+\mu_1\tau_2\nonumber\\
 &+E_3-\lambda_2\tau_3r_1\left(\frac{E_3}{\tau_3} \right)+\mu_1\tau_3+\frac {\kappa_uc_u^3l_u^3}{T^2}-\mu_2l_u\nonumber\\
 &+\left(\lambda_2+\lambda_3+\mu_1/f_{a,\max}-\mu_2\right)l_a-\mu_1T+\mu_2 L.\nonumber
\end{align}
Then the dual function of problem (P1.1) is
\begin{align}\label{eq.dual_function}
g( {\bm \lambda}, {\bm \mu}) = \min_{{\bm E},{\bm \tau}, {\bm l}} &~{{\cal L}}({\bm E},{\bm \tau}, {\bm l}, {\bm \lambda}, {\bm \mu})\\
{\rm s.t.}&~\eqref{eq:f0:max},~\eqref{eq:f1:max},~\eqref{eq.part_prob.c},~\eqref{eq.part_prob.d},~\eqref{eq.part_prob.1.e},~{\text{and}}~\eqref{eq.part_prob.1.e1}. \nonumber
 \end{align}
Consequently, the dual problem of (P1.1) is
\begin{align}
{\rm (D1.1)}:~\max_{{\bm \lambda}, {\bm \mu}} &\quad g({\bm \lambda},{\bm \mu})\label{eq.dual_prob}\\
{\rm s.t.}&~~ \mu_1 \geq 0,~\lambda_i\ge 0,~ \forall i\in\{1,2,3\}.\nonumber
\end{align}
We denote $\cal X$ as the set of $({\bm \lambda},{\bm \mu})$ characterized by the constraints in \eqref{eq.dual_prob}.

Since problem $({\rm P1.1})$ is convex and satisfies the Slater's condition, strong duality holds between problems $({\rm P1.1})$ and $({\rm D1.1})$. As a result, one can solve $({\rm P1.1})$ by equivalently solving its dual problem $({\rm D1.1})$. In the following, we first obtain the dual function $g({\bm \lambda},{\bm \mu})$ for any given $({\bm \lambda},{\bm \mu})\in \cal X$, and then obtain the optimal dual variables to maximize $g({\bm \lambda},{\bm \mu})$. For convenience of presentation, we denote $({\bm E}^*,{\bm \tau}^*,{\bm l}^*)$ as the optimal solution to (\ref{eq.dual_function}) under any given $({\bm \lambda},{\bm \mu})\in \cal X$, $({\bm E}^{\rm opt},{\bm \tau}^{\rm opt},{\bm l}^{\rm opt})$ as the optimal primal solution to $({\rm P1.1})$, and $({\bm \lambda}^{\rm opt},{\bm \mu}^{\rm opt})$ as the optimal dual solution to problem $({\rm D1.1})$.

\subsubsection{Derivation of Dual Function $g(\mv \lambda, \mv \mu)$} \label{sec:evalution}

First, we obtain $g(\mv \lambda, \mv \mu)$ by solving (\ref{eq.dual_function}) under any given  $(\mv \lambda,\mv \mu)\in{\cal X}$. Note that (\ref{eq.dual_function}) can be decomposed into the following five subproblems.
\begin{align}
&\min_{E_1, \tau_1, l_h} E_1+\mu_1\tau_1-\lambda_1\tau_1r_{01}\left(\frac{E_1}{\tau_1}\right)+\frac{\kappa_hc_h^3l_h^3}{(T-\tau_1)^2} +(\lambda_1-\mu_2)l_h \notag \\
& ~~{\rm s.t.}~~\eqref{eq:f1:max}~{\rm and}~0\leq E_1\leq \tau_1P_{u,\max}\notag \\
&~~~~~~~~~0\leq \tau_1\leq T,~l_h \ge 0. \label{eq.sub_part_prob1}
\end{align}
\begin{align}
&\min_{E_2, \tau_2 } ~ E_2+\mu_1\tau_2-\lambda_2\tau_2r_0\left(\frac{E_2}{\tau_2}\right)-\lambda_3\tau_2r_{01}\left(\frac{E_2}{\tau_2}\right) \notag \\
&~~{\rm s.t.}~~0\leq E_2\leq \tau_2P_{u,\max},~~0\leq \tau_2\leq T. \label{eq.sub_part_prob2}
\end{align}
\begin{align}
 &\min_{E_3, \tau_3} ~ E_3+\mu_1\tau_3-\lambda_2\tau_3r_{1}\left(\frac{E_3}{\tau_3}\right)  \notag \\
 &~~{\rm s.t.}~~0\leq E_3\leq \tau_3P_{h,\max},~~0\leq \tau_3\leq T.\label{eq.sub_prob3}
\end{align}
\begin{align} 
&\min_{l_u\ge 0}  ~~\frac{\kappa_uc_u^3l_u^3}{T^2}-\mu_2l_u \notag \\
&~~{\rm s.t.}~~ c_ul_u\leq Tf_{u,\max}.\label{eq.sub_prob4} 
\end{align}
\begin{align}
&\min_{0\leq l_a\leq L}  ~\left(\lambda_2+\lambda_3+\mu_1/f_{a,\max}-\mu_2\right)l_a. \label{eq.sub_prob5}
\end{align}
For problems \eqref{eq.sub_part_prob1}--\eqref{eq.sub_prob5}, we present their optimal solutions in the following lemmas. Due to the similar structures of problems  \eqref{eq.sub_part_prob1}--\eqref{eq.sub_prob3}, we present the proof of Lemma \ref{lem1} and omit the proofs of Lemmas \ref{lem2}--\ref{lem4} for brevity. 

\begin{lemma}\label{lem1}
Under given $( {\bm \lambda},{\bm \tau})\in{\cal X}$, the optimal solution $(E_1^*,\tau_1^*,l_h^*)$ to problem \eqref{eq.sub_part_prob1} satisfies
\begin{align}
&E_1^*=P_1^*\tau_1^*,\label{eqn:E1^*}\\
&l_h^*=M_1^*(T-\tau_1^*),\label{eqn:lh^*}\\
\label{eqn:tau1}
&\tau_1^*
\begin{cases}
=T,&{\rm if} ~\rho_1<0,\\
\in[0,T],&{\rm if}~\rho_1=0,\\
=0,&{\rm if} ~\rho_1> 0,
\end{cases}
\end{align}
where $P_1^*=\left[\frac{\lambda_1B}{\ln2}-\frac{\sigma_1^2}{h_{01}}\right]^{P_{u,\max}}_0$ with $[x]^a_b\triangleq\min\{a,\max\{x,b\}\}$ and
\begin{align}
M_1^*=
\begin{cases}
\left[\sqrt{\frac{\mu_2-\lambda_1}{3\kappa_hc_h^3}}\right]^{\frac{f_{h,\max}}{c_h}}_0 ,&{\rm if} ~ \mu_2-\lambda_1\geq 0,\\
0,&{\rm if} ~ \mu_2-\lambda_1<0,
\end{cases}
\end{align}
\begin{align}\label{eqn:rho1}
\rho_1 = &\mu_1-\lambda_1r_{01}(P_1^*)+2\kappa_h(c_hM_1^*)^3+\frac{\lambda_1BP_{1}^*{h_{01}}/{\sigma_1^2}}{(1+P_{1}^*{h_{01}}/{\sigma_1^2})\ln2} \nonumber \\
&-\alpha_1 P_{u,\max}+\frac{\beta_1f_{h,\max}}{c_h},
\end{align}
\begin{align}\label{eqn:part:sub1:a2}
\alpha_1=&
\begin{cases}
0, &{\rm if}~P_1^*<P_{u,\max},\\
\frac{\lambda_1B{h_{01}}/{\sigma_1^2}}{\ln2\left(1+P_1^*h_{01}/{\sigma_1^2}\right)}-1,&{\rm if}~P_1^*=P_{u,\max},
\end{cases}
\end{align}

\begin{align}\label{eqn:part:sub1:b2}
\beta_1=&
\begin{cases}
0, &{\rm if}~M_1^*<\frac{f_{h,\max}}{c_h},\\
\mu_2-\lambda_1-3\kappa_hc_h^3(M_1^*)^2,&{\rm if}~M_1^*=\frac{f_{h,\max}}{c_h}.
\end{cases}
\end{align}
\end{lemma}

\begin{IEEEproof}
See Appendix A.
\end{IEEEproof}


\begin{lemma}\label{lem2}
Under given $( {\bm \lambda},{\bm \tau})\in{\cal X}$, the optimal solution $(E_2^*,\tau_2^*)$ to problem (\ref{eq.sub_part_prob2}) satisfies
\begin{align}
&E_2^* = P_2^*\tau_2^*,\label{eqn:E2}\\
\label{eqn:tau2}
&\tau_2^*
\begin{cases}
=T,&{\rm if} ~\rho_2<0,\\
\in[0,T],&{\rm if}~\rho_2=0,\\
=0,&{\rm if} ~\rho_2> 0,
\end{cases}
\end{align}
where $P_2^*=\left[\frac{\sqrt{v^2-4uw}-v}{2u}\right]^{P_{u,\max}}_0$ with $u=\frac{\ln2}{B}\frac{h_{0}}{\sigma_0^2}\frac{h_{01}}{\sigma_1^2}$,
$v=\frac{\ln2}{B}(\frac{h_{0}}{\sigma_0^2}+\frac{h_{01}}{\sigma_1^2})-(\lambda_2+\lambda_3)\frac{h_{0}}{\sigma_0^2}\frac{h_{01}}{\sigma_1^2}$,
$w=\frac{\ln2}{B}-\lambda_2\frac{h_0}{\sigma_0^2}-\lambda_3\frac{h_{01}}{\sigma_1^2}$, $
\rho_2=\mu_1-\lambda_2r_{0}(P_2^*)+\frac{\lambda_2BP_2^*\frac{h_0}{\sigma_0^2}}{(1+P_2^*\frac{h_0}{\sigma_0^2})\ln2}-\lambda_3r_{01}(P_2^*) +\frac{\lambda_3BP_2^*\frac{h_{01}}{\sigma_1^2}}{(1+P_2^*\frac{h_{01}}{\sigma_1^2})\ln2}-\alpha_2P_{u,\max}$, and
\begin{align*}
\alpha_2=&
\begin{cases}
0, &{\rm if}~P_2^*<P_{u,\max},\\
\frac{\lambda_3B\frac{h_{01}}{\sigma_1^2}}{(1+P_2^*\frac{h_{01}}{\sigma_1^2})\ln2}+\frac{\lambda_2B\frac{h_0}{\sigma_0^2}}{(1+P_2^*\frac{h_0}{\sigma_0^2})\ln2}-1,&{\rm if}~P_2^*=P_{u,\max}.
\end{cases}
\end{align*}
\end{lemma}

\begin{lemma}\label{lem3}
Under given $( {\bm \lambda},{\bm \tau})\in{\cal X}$, the optimal solution $(E_3^*,\tau_3^*)$ to problem (\ref{eq.sub_prob3}) satisfies
\begin{align}
&E_3^* = P_3^*\tau_3^*,\label{eqn:E3}\\
\label{eqn:tau3}
&\tau_3^*
\begin{cases}
=T,&{\rm if} ~\rho_3<0,\\
\in[0,T],&{\rm if}~\rho_3=0,\\
=0,&{\rm if} ~\rho_3> 0,
\end{cases}
\end{align}
where $P_3^*=\left[\frac{\lambda_2B}{\ln2}-\frac{\sigma_1^2}{h_{1}}\right]^{P_{h,\max}}_0$ and $\rho_3=\mu_1+\frac{\lambda_2BP_3^*\frac{h_{1}}{\sigma_1^2}}{(1+P_3^*\frac{h_{1}}{\sigma_1^2})\ln2}-\lambda_2r_{1}(P_3^*)-\alpha_3P_{h,\max}$ with
\begin{align*}
\alpha_3=&
\begin{cases}
0, &{\rm if}~P_3^*<P_{h,\max},\\
\frac{\lambda_2B{h_{1}}/{\sigma_0^2}}{(1+P_3^*{h_{1}}/{\sigma_0^2})\ln2}-1,&{\rm if}~P_3^*=P_{h,\max}.
\end{cases}
\end{align*}
\end{lemma}

\begin{lemma}\label{lem4}
For given $( {\bm \lambda},{\bm \tau})\in{\cal X}$, the optimal solution $l_u^*$ to problem (\ref{eq.sub_prob4}) is
\begin{align}\label{eqn:l0}
l_u^*=\left[T\sqrt{\frac{\mu_2}{3\kappa_uc_u^3}} \right]^{\frac{Tf_{u,\max}}{c_u}}_0.
\end{align}
\end{lemma}

\begin{lemma}\label{lem5}
For given $( {\bm \lambda},{\bm \tau})\in{\cal X}$, the optimal solution $l_a^*$ to problem (\ref{eq.sub_prob5}) is
\begin{align}
l_a^*\left\{
\begin{array}{ll}
=0,&{\rm if}~\lambda_2+\lambda_3+\mu_1/f_{a,\max}-\mu_2>0,\\
\in [0,L],&{\rm if}~\lambda_2+\lambda_3+\mu_1/f_{a,\max}-\mu_2 = 0,\\
=L,&{\rm if}~ \lambda_2+\lambda_3+\mu_1/f_{a,\max}-\mu_2<0.
\end{array}\right.\label{eqn:l_a:star}
\end{align}
\end{lemma}

\begin{IEEEproof}
Note that the objective function is linear with respect to $l_a$ when $\lambda_2+\lambda_2+\mu_1/f_{a,\max}-\mu_2\neq 0$. The proof of Lemma \ref{lem5} is straightforward; we then omit it herein.
\end{IEEEproof}

Note that in \eqref{eqn:tau1}, \eqref{eqn:tau2}, \eqref{eqn:tau3}, or \eqref{eqn:l_a:star}, if $\rho_i=0$ (for any $i\in\{1,2,3\}$) or $\lambda_2+\lambda_3+\mu_1/f_{a,\max}-\mu_2=0$, then the optimal solution $\tau_i^*$ or $l_a^*$ is non-unique in general. In this case, we choose $\tau_i^* = 0$ and $l_a^*=0$ for the purpose of evaluating the dual function $g( {\bm \lambda}, {\bm \mu})$. It is worth noting that such choices may not be feasible nor optimal for the primal problem (P1.1). To tackle this issue, we will use an additional step in Section \ref{sec:find:primary} later to find the primal optimal $\tau_i^{\rm opt}$'s and $l_a^{\rm opt}$ for (P1.1).
%

By combining Lemmas~\ref{lem1}--\ref{lem5}, the dual function $g({\bm \lambda},{\bm \mu})$ is obtained for any given $({\bm \lambda},{\bm \mu})\in{\cal X}$.

\subsubsection{Obtaining $\mv \lambda^{\rm opt}$ and $\mv \mu^{\rm opt}$ to Maximize $g(\mv \lambda, \mv \mu)$}

Next, we search over $({\bm \lambda},{\bm \mu})\in{\cal X}$ to maximize $g(\mv \lambda, \mv \mu)$ for solving problem $({\rm D1.1})$. Since the dual function $g({\mv \lambda},{\mv \mu})$ is always concave but non-differentiable in general, we can use subgradient based methods, such as the ellipsoid method, to obtain the optimal $\mv \lambda^{\rm opt}$ and $\mv \mu^{\rm opt}$ for $({\rm D1.1})$.
Note that for the objective function in problem \eqref{eq.dual_function}, the subgradient with respect to $(\mv \lambda,\mv \mu)$ is
\begin{align*}
\left[ l_h^*-\tau_1^*r_{01}\left(\frac{E_1^*}{\tau_1^*}\right), l_a^*-\tau_2^*r_0\left(\frac{E_2^*}{\tau_2^*}\right)-\tau_3^*r_1\left(\frac{E_3^*}{\tau_3^*}\right), \right.\\
\left. l_a^*-\tau_2^*r_{01}\left(\frac{E_2^*}{\tau_2^*}\right),\sum_{i=1}^3 \tau_i^*+l_a^*/f_{a,\max}-T,L-l_u^*-l_h^*-l_a^*  \right].
\end{align*}
For the constraints $\mu_1\geq 0$ and $\lambda_i\geq 0$, the subgradients are $\mv e_4$ and $\mv e_i$, $\forall i\in\{1,2,3\}$, respectively, where $\boldsymbol{e}_i\in\mathbb{R}^5$ is the standard unit vector with one in the $i$-th entry and zeros elsewhere.

\subsubsection{Optimal Solution to $({\rm P1})$}\label{sec:find:primary}
With $\mv \lambda^{\rm opt}$ and $\mv \mu^{\rm opt}$ obtained, it remains to determine the optimal solution to problem $({\rm P1.1})$ (and thus $({\rm P1})$). By replacing $\mv \lambda$ and $\mv \mu$ in Lemmas \ref{lem1}--\ref{lem5} as $\mv \lambda^{\rm opt}$ and $\mv \mu^{\rm opt}$, we denote the corresponding $P_i^*$'s, $l_u^*$, and $M_1^*$ as $P_i^{\rm opt}$'s, $l_u^{\rm opt}$, and $M_1^{\rm opt}$, respectively. Accordingly, $\mv P^{\rm opt} = [P_1^{\rm opt},P_2^{\rm opt},P_3^{\rm opt}]$ corresponds to the optimal solution of $\mv P$ to problem $({\rm P1})$, and $l_u^{\rm opt}$ corresponds to the optimal solution of $l_u$ to both problems $({\rm P1})$ and $({\rm P1.1})$. Nevertheless, due to the non-uniqueness of $\tau_i^*$'s and $l_a^*$, we need an additional step to construct the optimal solution of other variables to problem $({\rm P1})$. Fortunately, with ${\mv P}^{\rm opt}$, $M_1^{\rm opt}$, and $l_u^{\rm opt}$ obtained, we know that the optimal solution must satisfy $l_h =M_1^{\rm opt}(T-\tau_1)$ and $E_i =P_i^{\rm opt} \tau_i, \forall i\in\{1,2,3\}$. By substituting them in $({\rm P1})$ or $({\rm P1.1})$, we have the following LP to obtain $\mv \tau^{\rm opt}$ and $l_a^{\rm opt}$.
\begin{align}\label{eq.part_prob.2}
\min_{\mv \tau, l_a\ge 0} ~& \sum_{i=1}^3 \tau_iP_i^{\rm opt}+{\kappa_h(c_hM_1^{\rm opt})^3(T-\tau_1)} \\ 
{\rm s.t.} ~&M_1^{\rm opt}(T-\tau_1)\leq \tau_1r_{01}(P_1^{\rm opt})\notag\\
~& l_a\leq \tau_2r_0(P_2^{\rm opt})+\tau_3r_1(P_3^{\rm opt})\notag \\
~& l_a \leq\tau_2r_{01}(P_2^{\rm opt})\notag \\
~&M_1^{\rm opt}(T-\tau_1) + l_a+ l_u^{\rm opt}= L\notag \\
~& \eqref{eqn:T1}~{\text{and}}~ 0\le \tau_i\le T,~\forall i\in\{1,2,3\}.\notag
\end{align}
The LP in (\ref{eq.part_prob.2}) can be efficiently solved by the standard interior-point method\cite{Boyd2004}. By combining $\mv \tau_p^{\rm opt}$, $l_h^{\rm opt}$, and $l_a^{\rm opt}$, together with ${\mv P}^{\rm opt}$ and $l_u^{\rm opt}$, the optimal solution to problem $({\rm P1})$ is finally found.

%

\section{Numerical Results}\label{sec:simulation}
In this section, we present numerical results to validate the performance of the proposed joint computation and communication cooperation design, as compared to the following benchmark schemes without such a joint design.
\begin{itemize}
  \item {\it Local computing}: the user executes the computation tasks locally by itself. The minimum energy consumption can be obtained as $E_u^{\rm loc}={\kappa_u c_u^3L^3}/{T^2}$.
  \item {\it Computation cooperation}: the computation tasks are partitioned into two parts for the user's local computing and offloading to the helper, respectively.  This corresponds to solving problem (P1) by setting $l_a=0$ and $\tau_2=\tau_3=0$.
  \item {\it Communication cooperation}: the computation tasks are partitioned into two parts for the user's local computing and offloading to the AP, respectively. The offloading is assisted by the helper's communication cooperation as a DF relay. This corresponds to solving problem (P1) by setting $l_h=0$ and $\tau_2+\tau_3=T$.
\end{itemize}

In the simulation, we consider that the user and the AP are located with a distance of $250$ meters (m) and the helper is located on the line between them. Let $D$ denote the distance between the user and the helper. The path-loss between any two nodes is denoted as $\beta_0\left( \frac{d}{d_0}\right)^{-\zeta}$, where $\beta_0=-60$ dB is the path loss at the reference distance of $d_0=10$ m, $d$ denotes the distance from the transmitter to the receiver, and $\zeta=3$ denotes the path-loss exponent.
Furthermore, we set $B=1$~MHz, $\sigma_0^2=\sigma_1^2=-70$~dBm, $c_u=c_h=10^3$ cycles/bit, $\kappa_u=10^{-27}$, $\kappa_h =0.3 \times 10^{-27}$, $P_{u,\max}=P_{h,\max}=40$~dBm, $f_{u,\max}=2$~GHz, $f_{h,\max}=3$~GHz, and $f_{a,\max}=5$~GHz.

\begin{figure}
\centering
    \includegraphics[width=9.0cm]{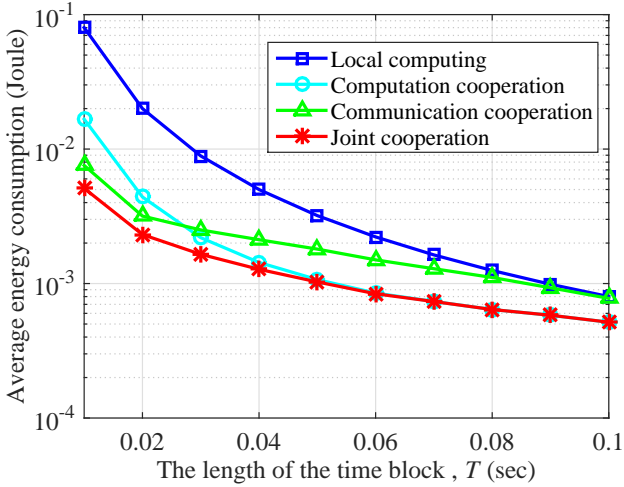}
\caption{The average energy consumption versus the time block length.} \label{fig_sim_T1}
\end{figure}

\begin{figure}
\centering
    \includegraphics[width=9.0cm]{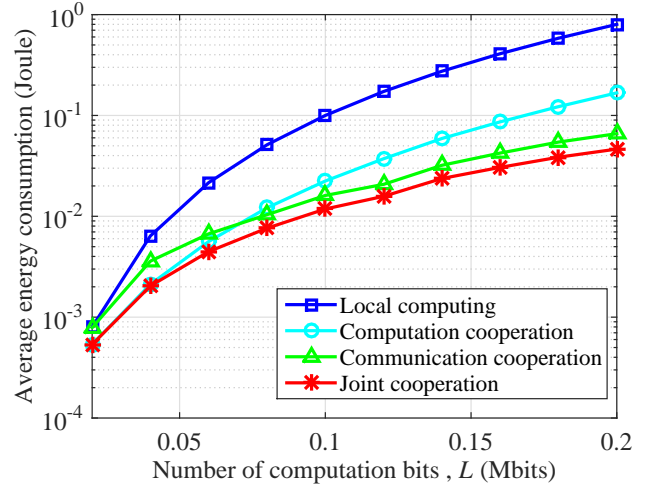}
\caption{The average energy consumption versus the number of computation bits.} \label{fig_sim_L1}
\end{figure}

Fig.~\ref{fig_sim_T1} shows the average energy consumption versus the time block length $T$, where $L=0.02$~Mbits and $D=120$~m. It is observed that the average energy consumption by all the schemes decreases as $T$ increases.
The communication-cooperation scheme is observed to achieve lower energy consumption than the computation-cooperation scheme when $T$ is small (e.g., $T<0.03$~sec); while the reverse is true when $T$ becomes large. It is also observed that the computation-cooperation and the communication-cooperation schemes both outperform the local-computing scheme, due to the fact that the two cooperation based schemes additionally exploit computation resources at the helper and the AP, respectively. The proposed joint-cooperation scheme is observed to achieve the lowest energy consumption.

Fig.~\ref{fig_sim_L1} depicts the average energy consumption versus the number of computation bits $L$, where $T=0.1$~sec and $D=120$~m. In general, similar observations are as in Fig.~\ref{fig_sim_T1}.
In particular, it is observed that at small $L$ values (e.g., $L < 0.06$ Mbits), the local-computing scheme achieves a similar performance as the joint-cooperation scheme. When $L$ becomes larger, the benefit of joint computation and communication cooperation is observed.

%
%

\section{Conclusion}\label{sec:conclusions}


In this paper, we investigated a new joint computation and communication cooperation approach in a simplified three-node MEC system, where a nearby helper node is enabled to share its computation and communication resources to help improve the user's performance for mobile computation. We proposed a four-slot protocol to enable this approach and developed a new energy-efficient design framework to minimize the total energy consumption at both the user and the helper while meeting the computation latency requirements, by jointly allocating their computation and communication resources. It is our hope that this paper can open a new avenue in exploring the multi-resource user cooperation to improve the computation performance for MEC.
\appendix
\subsection{Proof of Lemma~\ref{lem1}}
\vspace{-0cm}

As problem (\ref{eq.sub_part_prob1}) is convex and satisfies the Slater's condition, strong duality holds between problem (\ref{eq.sub_part_prob1}) and its dual problem. Therefore, one can solve this problem by applying the Karush-Kuhn-Tucker (KKT) conditions~\cite{Boyd2004}. The Lagrangian of problem (\ref{eq.sub_part_prob1}) is given by
\begin{align*}
{\cal L}_1=&E_1+\mu_1\tau_1-\lambda_1\tau_1r_{01}(\frac{E_1}{\tau_1})-\mu_2l_h+\lambda_1l_h+\frac{\kappa_hc^3l_h^3}{(T-\tau_1)^2}\\
&-a_1E_1+\alpha_1(E_1-\tau_1 P_{u,\max})-b_1\tau_1+b_2(\tau_1-T)\\
&-d_1l_h+\beta_1\left(l_h-\frac{(T-\tau_1)f_{h,\max}}{c}\right),
\end{align*}
where $a_1, \alpha , b_1, b_2$, $d_1$, and $\beta_1$ are the non-negative Lagrange multipliers associated with $E_1\geq0, E_1\leq \tau_1 P_{u,\max}, \tau_1\geq0, \tau_1\leq T$, $l_h\geq0$, and $l_h\leq\frac{(T-\tau_1)f_{h,\max}}{c}$, respectively.

Based on the KKT conditions, it follows that
\begin{subequations}\label{eq.kkt}
\begin{align}
&\textstyle a_1E_1=0, ~\alpha_1 (E_1-\tau_1 P_{u,\max})=0, ~b_2(\tau_1-T)=0\\
&\textstyle ~b_1\tau_1=0,~d_1l_h=0,~\beta_1\left(l_h-\frac{(T-\tau_1)f_{h,\max}}{c}\right)=0\\
&\textstyle \frac{\partial {\cal L}_1}{\partial E_1}=1-\frac{\lambda_1B\frac{h_{01}}{\sigma_1^2}}{\ln2\left(1+\frac{E_1}{\tau_1}\frac{h_{01}}{\sigma_1^2}\right)}-a_1+\alpha_1 =0\\
& \textstyle\frac{\partial {\cal L}_1}{\partial \tau_1}= \frac{2\kappa_hc^3l_h^3}{(T-\tau_1)^3}+\mu_1-\lambda_1B\log_2 \left( 1+\frac {E_1}{\tau_1}\frac{h_{01}}{\sigma_1^2} \right)+\frac{\beta_1 f_{h,\max}}{c}\notag \\
&~~~~~~~~\textstyle+\frac{\lambda_1B\frac{h_{01}}{\sigma^2}\frac{E_1}{\tau_1}}{\ln2\left(1+\frac{E_1}{\tau_1}\frac{h_{01}}{\sigma_1^2}\right)}-b_1+b_2+\alpha_1 P_{u,\max}=0\\
&\textstyle\frac{\partial {\cal L}_1}{\partial l_h}=\frac{3\kappa_hc^3l_h^2}{(T-\tau_1)^2}-\mu_2+\lambda_1-d_1+\beta_1=0,
\end{align}
\end{subequations}
where (\ref{eq.kkt}a) and (\ref{eq.kkt}b) denote the complementary slackness condition, and (\ref{eq.kkt}c), (\ref{eq.kkt}d) and (\ref{eq.kkt}e) are the first-order derivative conditions of ${\cal L}_1$ with respect to $E_1$, $\tau_1$, and $l_h$, respectively.
Based on the KKT conditions, (\ref{eqn:E1^*}) follows from (\ref{eq.kkt}c), and (\ref{eqn:lh^*}) holds due to (\ref{eq.kkt}e). Furthermore, based on (\ref{eq.kkt}c), (\ref{eq.kkt}d), and (\ref{eq.kkt}e) and with some manipulations, we have \eqref{eqn:part:sub1:a2} and \eqref{eqn:part:sub1:b2}.

Furthermore, by substituting (\ref{eqn:E1^*}) and (\ref{eqn:lh^*}) into (\ref{eq.kkt}d) and assuming $\rho_1=b_2-b_1$, we thus have $\rho_1$ in \eqref{eqn:rho1}. Hence, the optimal $\tau_1^*$ is given in (\ref{eqn:tau1}). Until now, this lemma is proved.

\end{document}